\documentclass[12pt]{article}

\usepackage{amsmath} 
\usepackage{amssymb}
\usepackage{color}
\usepackage{graphicx}
\usepackage{booktabs}
\usepackage{amsthm}
\usepackage{lmodern}
\usepackage[margin=1.2in]{geometry}
\usepackage{setspace}

\renewcommand{\epsilon}{\varepsilon}

\begin{document}
\text{}
\vspace{1cm}
\thispagestyle{plain}
\begin{center}
\Large {\bf  Nonlocality in Quantum Gravity and the Breakdown of Effective Field Theory}
\vspace{1.5cm}

\normalsize
{ \sc Nicolás Valdés-Meller}\footnote{{\fontfamily{cmtt}\selectfont nvaldes@mit.edu}}
\vspace{1cm}

\normalsize
{\it Perimeter Institute for Theoretical Physics \\ 31 Caroline St N, Waterloo, ON N2L 2Y5, Canada}
\vspace{0.5cm}

{\it University of Waterloo \\ 200 University Ave W, Waterloo, ON N2L 3G1, Canada}

\vspace{1.5cm}
{\bf Abstract} 
\end{center}

\begin{singlespace}
\noindent We argue that quantum gravity is nonlocal, first by recalling well-known arguments that support this idea and then by focusing on a point not usually emphasized: that making a conventional effective field theory (EFT) for quantum gravity is particularly difficult, and perhaps impossible in principle. This inability to realize an EFT comes down to the fact that gravity itself sets length scales for a problem: when integrating out degrees of freedom above some cutoff, the effective metric one uses will be different, which will itself re-define the cutoff.
We also point out that even if the previous problem is fixed, naïvely applying EFT in gravity can lead to problems -- we give a particular example in the case of black holes.
\end{singlespace}

\vfill

\noindent {\it Essay written for the Gravity Research Foundation 2021 Awards for Essays on Gravitation. Submitted March 31, 2021. Received honorable mention.}
\vspace{0.2cm}

\newpage

\section*{Motivation and building blocks}

Locality is a key ingredient of most relativistic quantum field theories. It can be formulated as the requirement that spacelike-separated operators commute, which ensures cluster decomposition (experiments done ``far away'' are uncorrelated) \cite{Weinberg}. To be blunt, it would be crazy for a theory to be nonlocal -- fortunately the standard model of particle physics is local. Then there is a natural question we can ask: do we expect quantum gravity to be a local theory?

Here we argue that quantum gravity is likely not local. There is evidence pointing to this from various directions, some of which we mention below. However, we focus on issues that are not typically discussed. Namely, we associate the notion of locality to the possibility of properly setting up an effective field theory (EFT), and then claim that standard EFT ideas break down for the case of gravity.

EFT represents the principle that we can study the universe by separating it into different regimes, and using physical theories that are appropriate for each regime they seek to describe. 
For example, in particle physics the regimes are characterized by {\it length scales}, or alternatively energy scales. In quantum gravity the same assumption is often made  \cite{AS}.
The failure of this paradigm of separation of regimes would be just as crazy as the absence of locality.

One connection between locality and EFT is as follows: when theories are local we expect there to be a separation of length scales, which is in part what gives meaning to the concept of ``experiments that are far away'' in the first place. This separation of length scales is precisely what allows us to set up an EFT and separate physical processes into those length scales that we can probe, and those that we cannot \cite{EFT}. In this essay we emphasize one basic fact. In gravity, {\it the dynamical fields themselves set the length scales}, which precludes the starting point of the standard EFT paradigm. This is closely related to the issue that gravity is a background-independent theory \cite{Rovelli}; there is no background relative to which one can define long and short distances.

The ideas here are not stated rigorously: the focus is on intuition about the potentially relevant physics. This essay can be read as a collection of reflections and questions from a confused young theoretical physicist to his elders. Because essentially that is what it is.

\section*{Evidence for nonlocality}

Compelling evidence for nonlocality in quantum gravity comes from the construction of diffeomorphism-invariant operators, which requires a nonlocal dependence of observables on the metric. For instance, 
\cite{Dressing} found that even for operators in a flat spacetime background, when the operators were made to be gauge invariant in the presence of metric perturbations (``gravitationally dressed''), most of them acquired a dependence on the asymptotic metric.

Further clues come from string theory and holography. In string theory, there is the fact that string interactions are not pointwise but rather extended in space \cite{ST}. Additionally in holographic theories (whose main examples come from strings), there is a clear violation of locality, since the degrees of freedom in a region can be described by degrees of freedom on its boundary.
There is also circumstantial evidence that quantum gravity should be nonlocal; for instance, a unitary resolution of the black hole information problem can be achieved by abandoning absolute locality \cite{BH}.

\section*{Problems with EFT in gravity}

We proceed by recalling the Wilsonian perspective on EFT.
We start out with the path integral of a garden-variety quantum field theory 
\begin{align}
Z = \int \mathcal{D} \phi(x) e^{iS[\phi(x)]},
\end{align}
where $\phi$ are the set of fields in our theory and $S[\phi]$ is an integral over spacetime of local fields. We can also write our path integral by taking the Fourier transform of the fields $\phi(x)$ into $\varphi(k)$. Then we split our fields into ``UV'' and ``IR'' components such that $\varphi_{\text{UV}}(k)=0$ for $k<\Lambda$ and $\varphi_{\text{UV}}(k)=\varphi(k)$ for $k\geq \Lambda$ where $\Lambda$ is some energy scale (and the analogous definitions are made for $\varphi_{\text{IR}}$). Finally we ``integrate out'' UV modes in order to arrive at an effective theory which only probes long-distance degrees of freedom. More explicitly, 
\begin{align}
Z &= \int \mathcal{D} \varphi_{\text{IR}}(k) \int \mathcal{D} \varphi_{\text{UV}}(k) e^{iS[\varphi_{\text{UV}}+\varphi_{\text{IR}}]}  \nonumber \\
&\equiv \int^\Lambda \mathcal{D} \varphi_{\text{IR}}(k) e^{iS'[\varphi_{\text{IR}}]},
\end{align}
where $S'[\varphi_{\text{IR}}]$ is an action which in general will be nonlocal, but we replace the nonlocal interactions with local ones whose coefficients reflect how the UV fields affect IR physics \cite{Georgi}. This modifies UV behavior, but we don't mind because we only care about low-energy physics: we can still compute correlation functions of IR fields, or alternatively correlation functions of spacetime fields that are separated by distances $\gtrsim 1/\Lambda$. 

In attempting to extend this to quantum gravity we encounter several subtleties, but some of them are only technical. To name one: Fourier transforms aren't defined on general spacetimes, but this has been dealt with. For example, one can use the Laplacian associated to a metric to construct an analog of Fourier modes \cite{Reuter1}. The way we see it, there are two deeper (related) difficulties that arise in trying to define a gravitational EFT.

\subsection*{First problem: background dependence}

For quantum gravity, we expect the nonperturbative theory to be defined by a path integral over geometries.
One way of approaching this is to set up a background metric $\bar{g}_{\mu\nu}$, and integrate over metrics of the form $\bar{g}_{\mu\nu}+h_{\mu\nu}$ (taking care of technical details like integrating over ghost fields as well) \cite{Reuter1}. The purpose of $\bar{g}$ is to define what one means by ``IR" and ``UV"; this is the metric that sets length or energy scales when one wants to integrate out modes of different energies. Schematically, for Euclidean gravity the effective theory valid down to some IR scale $k$ would be described by
\begin{align}
Z = \int \mathcal{D} h_{\mu\nu} e^{-S[\bar{g}+h] - \Delta_k S[h;\bar{g}]} 
\end{align}
where $\Delta_k S$ is a nonlocal term, just like we would have for a normal EFT. Its purpose is to discriminate between high-energy and low-energy modes. However, it is of a more inadmissible nature here, because if we wish to integrate over {\it all background metrics}, getting rid of the nonlocal term by standard scaling arguments will not be possible -- this nonlocal operator itself depends on the background metric, e.g., containing terms of the form $(1/\bar{\nabla}^2)$, with $\bar{\nabla}$ defined in the standard way from $\bar{g}$.\footnote{In \cite{Reuter2}  the shortcoming of background dependence was acknowledged and partially addressed.}

As an example of how we could run into trouble, imagine starting out with a background metric and suggesting that we integrate out degrees of freedom above the energy scale $\Lambda$. Then we would have a new metric, which would re-define our cutoff scale to $\Lambda'$. If $\Lambda'>\Lambda$, we would be working with a theory which ignored degrees of freedom above $\Lambda$, but now claims to be capable of predictive power for even higher energies! The notion of {\it relative locality} was proposed in \cite{Relative} based on similar grounds.

\subsection*{Second problem: overconfidence in EFT}

Now let's imagine an observer who uses Einstein gravity and agrees that it is an EFT valid up to energy scales $\Lambda$. They are in flat space initially, and call their friends at infinity to set up a particle collision at energy $\Lambda$. The particles collide to form a black hole, which then proceeds to evaporate. Looking at the black hole from infinitely far away, our observer will see redshifted Hawking quanta (which had energies much greater than $\Lambda$ near the horizon) at energies below $\Lambda$. Thus they will treat these quanta with as much confidence as they would any other low-energy quantities. But these modes reflect UV degrees of freedom in the theory, inaccessible to the effective theory: using an effective field theory in gravity leads one to overestimate how effective the EFT is. 

This could be thought of as the trans-Planckian problem in disguise, but it suggests the potential for a more extreme violation of EFT, at scales much greater than the Planck length. The general lesson to take away is that our observer's overconfidence in EFT (trusting it to give information about degrees of freedom it shouldn't know about) is indicative of a more generic feature we could expect in quantum gravity, potentially originating from nonlocality. In studying black hole paradoxes we sometimes forget to extract the important lessons from the paradoxes -- we believe the important  thing to extract here is that EFT in gravity can lead us to make spurious conclusions about phenomena because we are overestimating the regime of validity of the EFT.

\section*{Concluding remarks}

Before wrapping up, we emphasize that an important question to keep in mind is: how badly is locality violated, and under which circumstances is gravity approximately local? This is a necessary issue to study if one wants to understand how our macroscopic world, which is local, can emerge from quantum gravity. Another thing to remember is that regimes are not characterized by length scales in all approaches to quantum gravity \cite{Spin}, so the above problems don't hold universally. 

On a more speculative note, perhaps the fundamental inability to separate length scales in quantum gravity is at the basis of why in quantum gravity we encounter essentially all areas of physics, regardless of regime of validity. Even if it is not a theory of everything, quantum gravity insists on bringing all of physics to the party.

\section*{Acknowledgements}

I wish to thank Bianca Dittrich, William Donnelly, Jake Hauser, and Sydney Timmerman for helpful discussions. I'm also grateful to Daniel Harlow and Jake Hauser for comments on the manuscript. Research at Perimeter Institute is supported in part by the Government of Canada through the Department of Innovation, Science
and Industry of Canada and by the Province of Ontario through the Ministry of Colleges and
Universities.


\singlespace


\begin{thebibliography}{9}
\bibitem{Weinberg} 
S. Weinberg,  ``The Quantum Theory of Fields". Cambridge: Cambridge University Press (1995). https://doi.org/10.1017/CBO9781139644167

\bibitem{AS} 
C.P. Burgess, ``Quantum Gravity in Everyday Life: General Relativity as an Effective Field Theory". Living Rev. Relativ. 7, 5 (2004). https://doi.org/10.12942/lrr-2004-5

\bibitem{EFT} 
A. Manohar, ``Introduction to Effective Field Theories'' (2017).
https://doi.org/10.1093/oso/9780198855743.001.0001. [hep-ph] 1804.05863.

\bibitem{Rovelli} 
C. Rovelli, ``Quantum Gravity". Cambridge: Cambridge University Press (2004). https://doi.org/10.1017/CBO9780511755804 

\bibitem{Dressing}
W. Donnelly and S. Giddings, ``Observables, gravitational dressing, and obstructions to locality and subsystems". Phys. Rev. D 94, 104038 (2016). https://doi.org/10.1103/PhysRevD.94.104038. [hep-th] 1607.01025. 

\bibitem{ST} 
G. Calcagni and L. Modesto, ``Nonlocality in string theory''. J. Phys. A: Math. Theor. 47 (2014) 355402. [hep-th] 1310.4957. 

\bibitem{BH}
S. Giddings, ``Nonlocality versus complementarity: a conservative approach to the information problem". Class. Quantum Grav. 28 025002 (2010). [hep-th] 0911.3395. 

\bibitem{Georgi} 
H. Georgi, ``Effective Field Theory''. Annual Review of Nuclear and Particle Science (1993) 43:1, 209-252. https://doi.org/10.1146/annurev.ns.43.120193.001233.


\bibitem{Reuter1}
M. Reuter, ``Nonperturbative evolution equation for quantum gravity". Phys. Rev. D 57, 971 (1998). https://doi.org/10.1103/PhysRevD.57.971. [hep-th] 9605030.


\bibitem{Reuter2}
C. Pagani and M. Reuter, ``Background independent quantum field theory and gravitating vacuum fluctuations". Annals of Physics Volume 411 (2019). https://doi.org/10.1016/j.aop.2019.167972. [gr-qc] 1906.02507.  


\bibitem{Relative}
G. Amelino-Camelia, L. Freidel, J. Kowalski-Glikman, and L. Smolin, ``Principle of relative locality". Phys. Rev. D 84, 084010 (2011). https://doi.org/10.1103/PhysRevD.84.084010. [hep-th] 1101.0931. 

\bibitem{Spin}
S. Steinhaus, ``Coarse Graining Spin Foam Quantum Gravity--A Review". Frontiers in Physics Volume 8 (2020). https://doi.org/10.3389/fphy.2020.00295. [gr-qc] 2007.01315. 



\end{thebibliography}
\end{document}